\documentclass[11pt,a4,dvips]{article}
\usepackage{moriond,epsfig}

\def\jpsi{\ensuremath{\mathrm{J}/\psi}}
\def\chic{\ensuremath{\chi_{\mathrm{c}}}}
\def\pt{\ensuremath{p_t}}
\def\antibar#1{\ensuremath{#1\bar{#1}}}%
\def\ppbar{\antibar{\mathrm{p}}}%
\def\epem{\ensuremath{\mathrm{e^+e^-}}}%
\def\pb{\mbox{pb$^{-1}$}}%
\def\GeV{\ifmmode {\mathrm{\ Ge\kern -0.1em V}}\else
                   \textrm{Ge\kern -0.1em V}\fi}%
\def\TeV{\ifmmode {\mathrm{\ Te\kern -0.1em V}}\else
                   \textrm{Te\kern -0.1em V}\fi}%

\begin{document}
\vspace*{4cm}
\title{B PHYSICS AT D\O}

\author{ F. FILTHAUT\\ (for the D\O\ Collaboration)}

\address{Department of Experimental High Energy Physics, Toernooiveld 1,\\
  6525 ED Nijmegen, The Netherlands}

\maketitle\abstracts{
  The D\O\ detector at FNAL has undergone a significant upgrade for its
  present Run II data taking period, allowing for a broad B physics programme.
  This report focuses on studies performed on a sample of approximately 75$\cdot$
  10$^{3}$ inclusive \jpsi\ events. Preliminary results on mass and lifetime
  measurements and prospects for flavour tagging are presented.}

\section{Introduction}

The high integrated luminosity expected to be delivered at Run II make the
Tevatron \ppbar\ collider an ideal environment for the study of a wide range of
physics processes. In particular, the high inclusive b production cross section
allows a precise study of many relatively rare B hadron decay
modes. The copious production of B$_{\mathrm{s}}$ mesons opens up a rich field of
B physics studies not accessible at \epem\ B factories. 
As presented previously in this meeting series~\cite{ref:verzocchi}, the D\O\
detector has been upgraded significantly to take advantage of these physics
opportunities.


\section{Data sample}
\label{sec:sample}

\begin{figure}[htb]
  \begin{center}
    \includegraphics[width=0.55\textwidth]{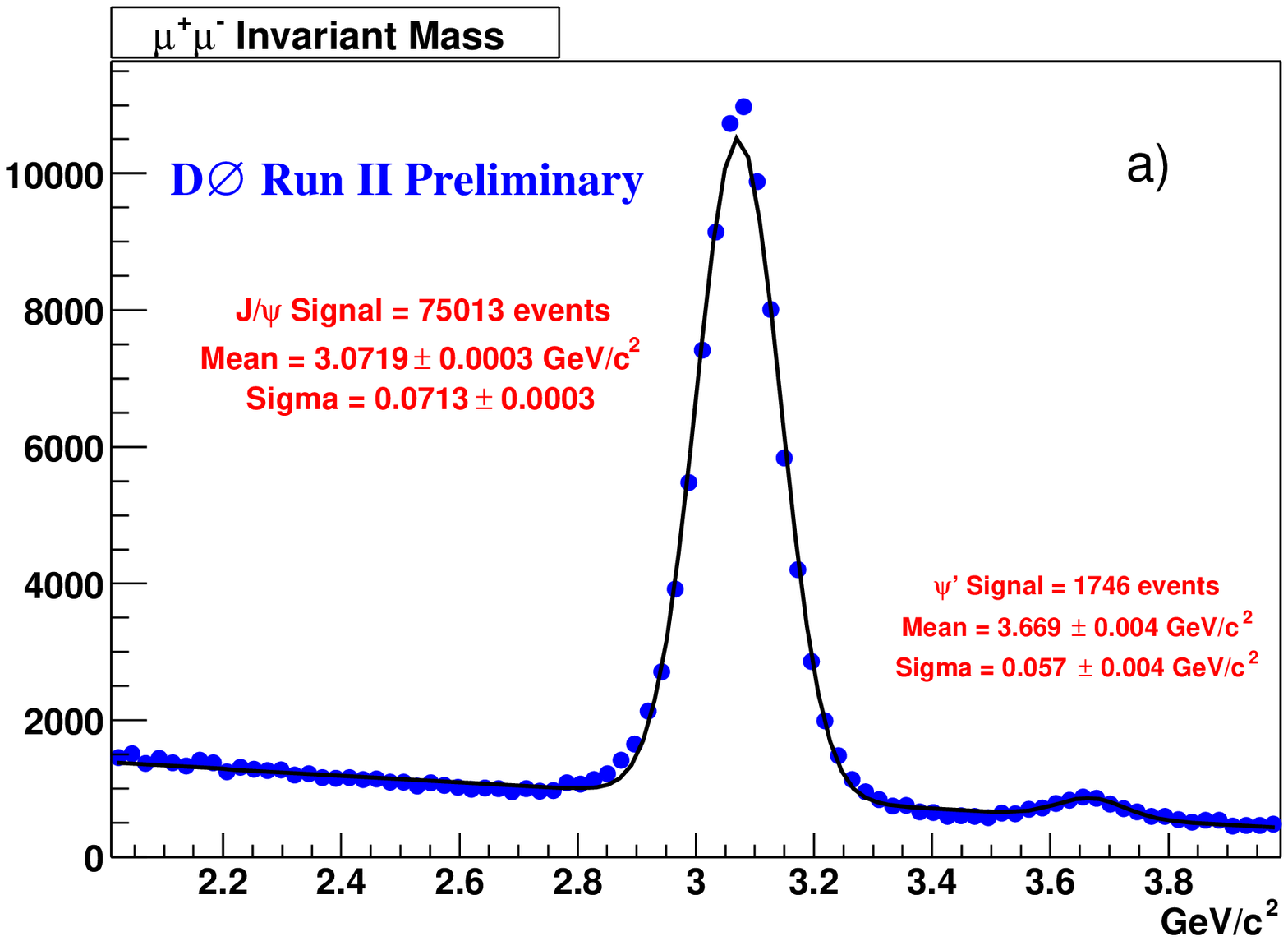}
    \includegraphics[width=0.39\textwidth]{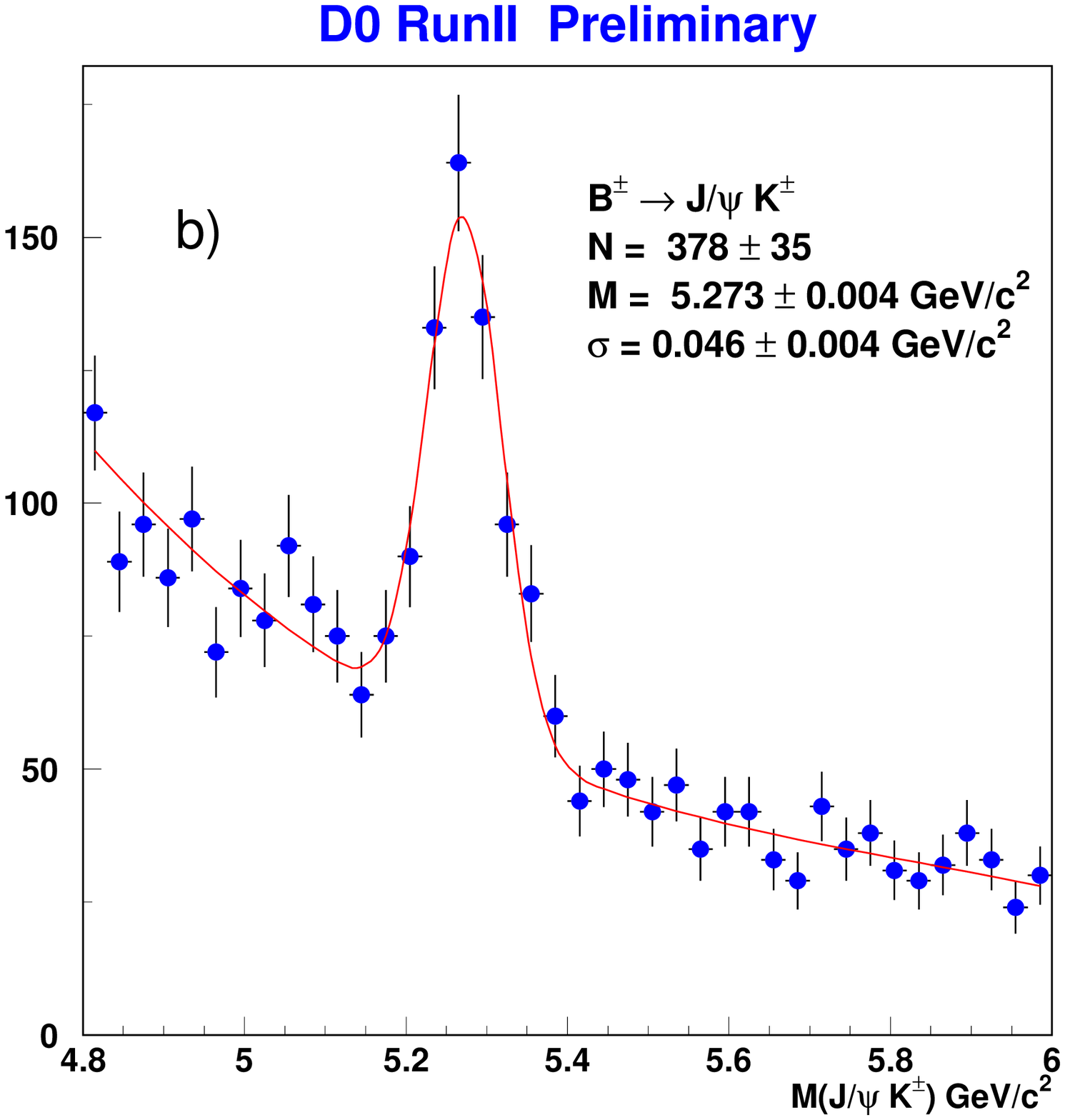}
    \caption{Invariant mass distribution of pairs of oppositely charged muons (a)
      and of $\jpsi \mathrm{K^{\pm}}$ candidates (b).}
    \label{fig:mmumu}
  \end{center}
\end{figure}

The data presented here represent an integrated luminosity of approximately 40\pb,
and were collected using di-muon triggers and pre-selected requiring two muons
reconstructed in the muon system within $|\eta| < 2.4$. The invariant mass
distribution of pairs of oppositely charged muons, as shown in Fig.~\ref{fig:mmumu},
indicates approximately 75$\cdot$10$^{3}$ inclusive
$\jpsi\rightarrow\mu^{+}\mu^{-}$
events. Given the preliminary calibration of the detector, the average observed
mass is in reasonable agreement with the nominal \jpsi\ mass~\cite{ref:pdg}.
Using this \jpsi\ sample, various exclusive B decays ($\mathrm{B^{\pm}\rightarrow
  \jpsi K^{(\ast)\pm}}$, $\mathrm{B_{d}\rightarrow\jpsi K^{\ast 0}}$, and
$\mathrm{B_{d}\rightarrow\jpsi K_{S}}$) 
have been identified. The mass distribution of $\jpsi\mathrm{K^{\pm}}$ events,
used extensively later in this report, is also shown in
Fig.~\ref{fig:mmumu}.
\vspace*{-2mm}

\section{\boldmath $\chic$ production}
\label{sec:chic}

The collected sample of inclusive \jpsi\ events is used to study \chic\ production
through its radiative decay $\chic\rightarrow\jpsi\gamma$. As low energy photons
are not easily identified in the D\O\ calorimeter, they are instead identified by
their photoconversions in the beam pipe and (mainly) in the detector materials.
With a kinematic cut $p_{\mathrm{T},\gamma} > 1 \GeV/c$, from Monte Carlo
computations the photon reconstruction efficiency is estimated to be approximately
0.37\%.

The photons thus reconstructed are combined with the found \jpsi\ candidates.
The mass difference resolution is insufficient to resolve the
individual \chic\ resonances, and in the fit the difference between the
$\chi_{\mathrm{c1}}$ and $\chi_{\mathrm{c2}}$ masses is therefore fixed to its
nominal value while their relative contribution to the peak is allowed to float
(the $\chi_{\mathrm{c0}}\rightarrow\jpsi\gamma$ branching ratio is too small to
contribute appreciably to the observed peak).
The resulting number of candidates, corrected for the photon reconstruction
efficiency and compared with the number of \jpsi\ candidates, leads to an
estimated fraction $F_{\chi}^{\jpsi}$ of \jpsi\ originating from \chic\ decays
\begin{equation}
  \label{eq:fchi}
  F_{\chi}^{\jpsi} = 0.030 \pm 0.04 (\mbox{stat.}),
\end{equation}
in good agreement with the Run I result from the CDF
Collaboration~\cite{ref:fchi-cdf} and in marked disagreement with predictions from
colour singlet models.

\section{B Lifetime}
\label{sec:blife}

The inclusive \jpsi\ events are used in two ways to perform B lifetime
measurements: once in an inclusive measurement, and once in an exclusive
$\mathrm{B^{\pm}\rightarrow\jpsi K^{\pm}}$ measurement. In both cases, the decay
lengths are estimated from the measured secondary vertex coordinates in the plane
transverse to the beam line, corrected using the B hadron \pt\ through
$(ct)_{\mathrm{B}} = \lambda_{\mathrm{B}} = L_{xy}
m_{\mathrm{B}}/\pt^{\mathrm{B}}$.

\subsection{Inclusive measurement}
\label{sec:blife-inc}

\begin{figure}[htb]
  \begin{center}
    \includegraphics[width=0.47\textwidth]{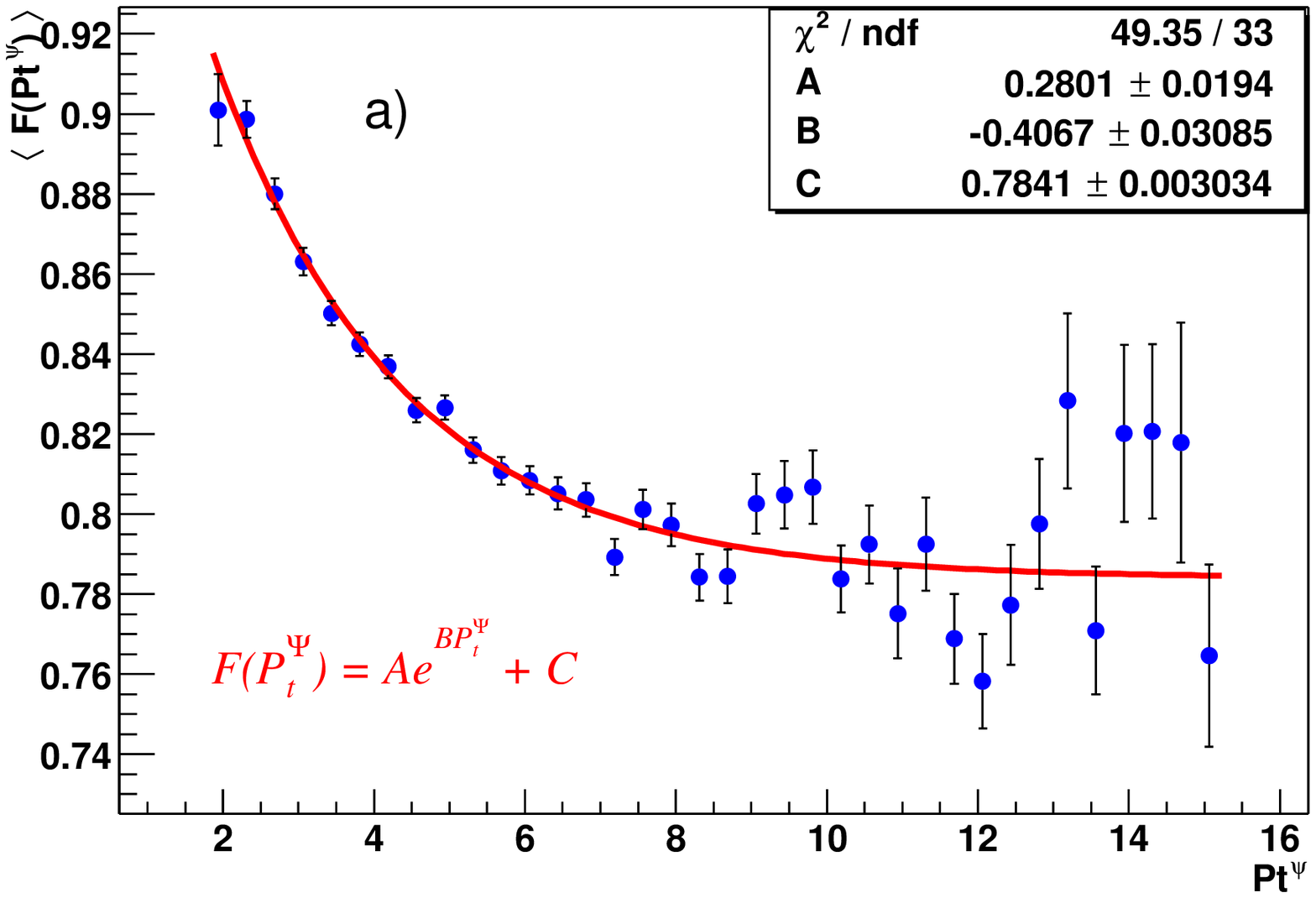}
    \includegraphics[width=0.47\textwidth]{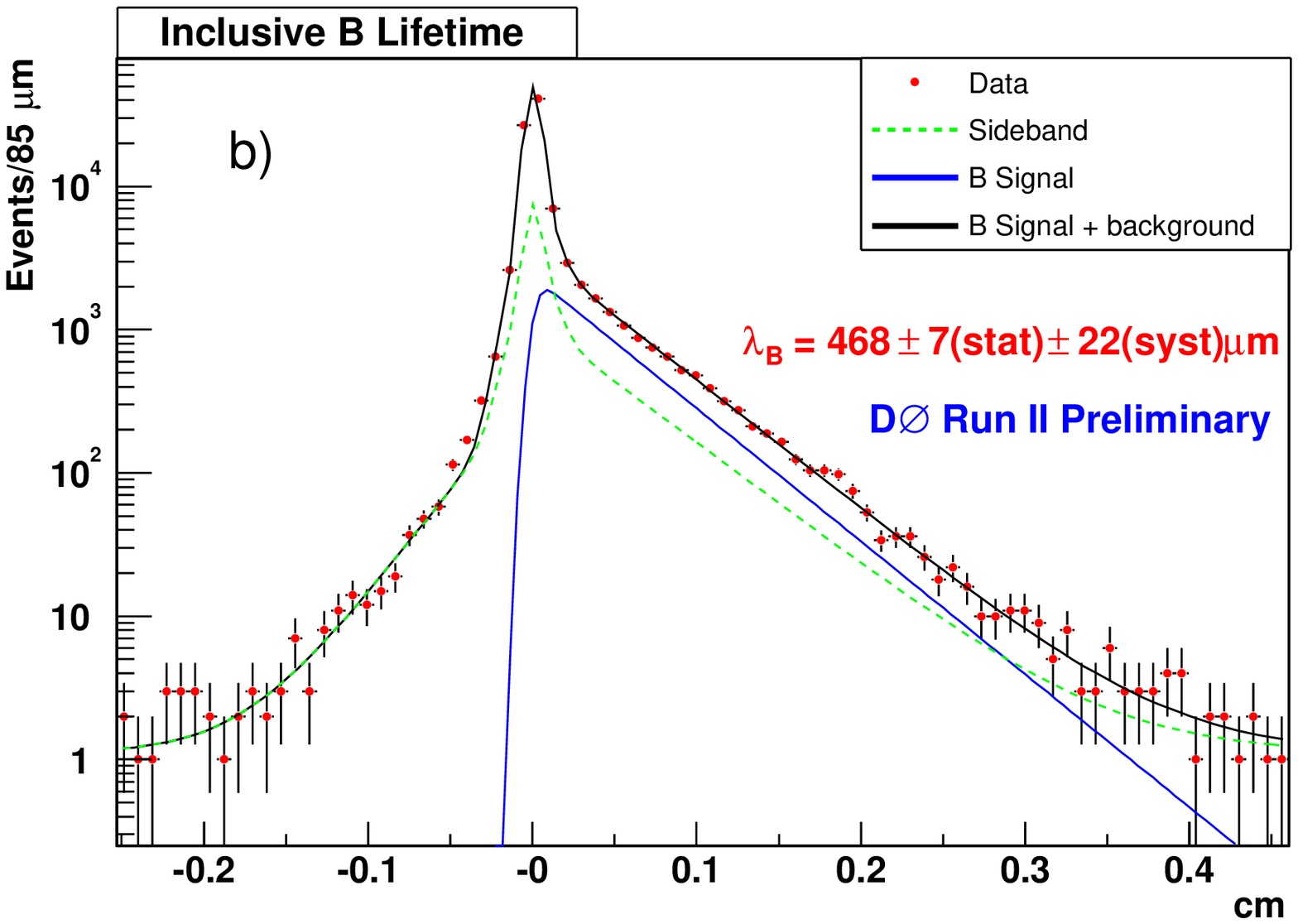}
    \caption{\jpsi\ \pt\ correction factor in the inclusive \jpsi\ lifetime
      analysis (a), and proper decay length distribution(b).}
    \label{fig:blife-inc}
  \end{center}
\end{figure}

In the inclusive measurement, the B hadron is not fully reconstructed so
that the quantity $m_{\mathrm{B}}/\pt^{\mathrm{B}}$ is not known from event to
event. It is therefore obtained as an average correction factor, as a function of
the \pt\ of the \jpsi. The correction factor is computed using the
Pythia Monte Carlo generator~\cite{ref:pythia}, with the B hadron decays simulated
by the QQ program~\cite{ref:QQ}, and is displayed in Fig.~\ref{fig:blife-inc}(a).
The proper decay length distribution of the \jpsi\ background is obtained from the
\jpsi\ sideband windows ($2.6\GeV < m_{\mu\mu} < 2.85\GeV$ and
$3.29\GeV < m_{\mu\mu} < 3.5\GeV$), and fit to a double Gaussian resolution
function and the sum of an exponential plus a constant background. This background
is fixed in the \jpsi\ signal window, and a prompt component and signal
exponential convoluted with the resolution function obtained from the sideband
windows is added. The signal window decay length distribution, along with its fit,
is shown in Fig.~\ref{fig:blife-inc}(b). The systematic uncertainty on the average
lifetime is dominated by uncertainties on the correction factor applied (16
$\mu$m, mainly from fragmentation uncertainties) and fit biases (13 $\mu$m, as
determined from Monte Carlo studies). The average lifetime is therefore determined
to be
\begin{equation}
  \label{eq:blife-inc}
  \tau_{\mathrm{B}} = (1.561 \pm 0.023 \mbox{(stat.)} \pm 0.073 \mbox{(syst.)})
  \mbox{ps},
\end{equation}
in good agreement with the world average result.

\subsection{$\mathrm{B^{\pm}}$ lifetime}
\label{sec:blife-bplus}

The use of fully reconstructed B hadrons allows to estimate the B hadron momentum
on an event by event basis and has lower background, and therefore leads to a much
reduced systematic lifetime uncertainty. A sample of $\mathrm{B^{\pm}}$ mesons is
obtained through their decay $\mathrm{B}^{\pm}\rightarrow\jpsi\mathrm{K^{\pm}}$,
where (in the absence of $\pi$/K separation) charged particles
compatible with originating from the \jpsi\ vertex are assigned the kaon mass.

The background to the lifetime distribution in this case consists of two
components: incompletely reconstructed B decays mainly populating the lower
$\mathrm{B^{\pm}}$ sideband region, and prompt combinatorial background. The
lifetime of the former is determined and fixed in the fit to the signal region,
while its fraction is fixed to expectations from Monte Carlo.
The $\mathrm{B^{\pm}}$ signal window is fit to this background plus an
exponential signal component convoluted with a Gaussian resolution function.
The result of the fit is
\begin{equation}
  \label{eq:blife-bplus}
  \tau_{\mathrm{B}^{\pm}} = (1.761 \pm 0.24 (\mbox{stat.})) \mbox{ps},
\end{equation}
within its large uncertainty again in agreement with the world average result.

\section{Flavour tagging}
\label{sec:flavour}

The measurement of mixing and CP violation in neutral B meson decays, a crucial
part of the Tevatron B physics programme, involves tagging the B meson's flavour
at its time of production. The use of reconstructed $\mathrm{B^{\pm}}$ mesons
allows to determine the performance of such flavour tags in an unbiased way.

In opposite side tags, the charge of the B meson of interest at its time of
production is estimated by considering the charge of the B hadron produced in
association with it.
Two such tags have been studied within the D\O\ Collaboration. In the soft muon
tag, the charge of the highest \pt\ muon (with $\pt > 1.9 \GeV/c$) in the event is
considered, requiring a separation from the signal B meson $\Delta R > 2$.
The jet charge tag considers the \pt\ weighted charge average $Q$ of the tracks
separated from the B meson in azimuth by $|\Delta\phi| > 2$ and associated with
the primary vertex within 2 cm. A jet is considered tagged if $|Q| > 0.2$.

The performance of the tags is characterised by factor $\epsilon D^{2}$, where
$\epsilon$ is the tagging efficiency and the ``dilution'' $D$ is given by
$D = 1 - 2P_{\mathrm{w}}$, with $P_{\mathrm{w}}$ being the fraction of wrongly
tagged jets. The ``raw'' efficiency and dilution, as obtained from the
$\mathrm{B^{\pm}}$ signal region, are corrected for the background under the
$\mathrm{B^{\pm}}$ peak using the corresponding numbers as obtained from the
$\mathrm{B^{\pm}}$ sideband windows. The corrected results are shown in
Table~\ref{table:flavour} and are in reasonable agreement with
expectations~\cite{ref:workshop}.
\begin{table}[htb]
  \begin{center}
    \begin{tabular}{|lr@{$\pm$}lr@{$\pm$}l|}
      \hline
      Tag & \multicolumn{2}{c}{soft muon} & \multicolumn{2}{c|}{jet charge}\\
      \hline
      efficiency $\epsilon$ (\%) &  8.2 &  2.2 & 55.1 &  4.1 \\
      dilution $D$ (\%)          & 63.9 & 30.1 & 21.1 & 10.6 \\
      $\epsilon D^{2}$ (\%)      &  3.3 &  1.8 &  2.4 &  1.7 \\
      \hline
    \end{tabular}
    \caption{Flavour tagging performance estimates.}
    \label{table:flavour}
  \end{center}
\end{table}
\vspace*{-4mm}

\section*{Conclusions}

Using a large sample of inclusive \jpsi\ events collected from an integrated
luminosity of about 40 \pb, the D\O\ Collaboration has 
searched for resonances decaying to $\jpsi X$. Preliminary B hadron lifetime
measurements are in reasonable agreement with world average
results. $\mathrm{B}^{\pm}\rightarrow\jpsi \mathrm{K^{\pm}}$ decays have been
used to demonstrate that D\O's flavour tagging performance is in agreement with
predictions made before the start of the Tevatron Run II.

Improvements are ongoing in the D\O\ detector performance, data collection
efficiency, and data reconstruction software programs. Together with the
increasing luminosity of the Tevatron collider, substantial improvements in the
results presented here are foreseen, as well as many qualitatively new results.

\section*{Acknowledgements}

It is a pleasure to acknowledge the help of my colleagues in the D\O\
Collaboration in preparing this report.

\section*{References}

\bibliography{moriond}


\end{document}